\documentclass[10pt]{article}

\usepackage{PRIMEarxiv}

\usepackage{algorithm}
\usepackage{algpseudocode}
\usepackage{amsmath}\usepackage{booktabs}
\usepackage{subcaption}
\usepackage{multirow}
\usepackage{bbding}
\usepackage{graphicx}       
\usepackage[backend=biber,style=ieee,sorting=none,natbib=true]{biblatex}
\usepackage{authblk}

\graphicspath{{media/}}     
\usepackage{bm}

\title{A Local Perspective-based Model for Overlapping Community Detection
}
\author[1,*]{Gaofeng Zhou}
\author[2,*,$\dagger$]{Rui-Feng Wang}
\author[3]{Kangning Cui}

\affil[1]{College of Computer Science \& Technology, Qingdao University, Qingdao 266071, China}
\affil[2]{College of Engineering, China Agricultural University, 17 Qinghua East Road, Haidian, Beijing 100083, China}
\affil[3]{Department of Mathematics, City University of Hong Kong, Kowloon, Hong Kong SAR, China}

\begin{document}
\maketitle

\let\thefootnote\relax
\footnotetext{\noindent $*$ Gaofeng Zhou and Rui-Feng Wang contributed equally to this work. Correspondence: sweefongreggiewong@cau.edu.cn}

\begin{abstract}

Community detection, which identifies densely connected node clusters with sparse between-group links, is vital for analyzing network structure and function in real-world systems. Most existing community detection methods based on GCNs primarily focus on node-level information while overlooking community-level features, leading to performance limitations on large-scale networks. To address this issue, we propose LQ-GCN, an overlapping community detection model from a local community perspective. LQ-GCN employs a Bernoulli-Poisson model to construct a community affiliation matrix and form an end-to-end detection framework. By adopting local modularity as the objective function, the model incorporates local community information to enhance the quality and accuracy of clustering results. Additionally, the conventional GCNs' architecture is optimized to improve the model’s capability in identifying overlapping communities in large-scale networks. Experimental results demonstrate that LQ-GCN achieves up to a 33\%  improvement in Normalized Mutual Information (NMI)  and a 26.3\% improvement in Recall compared to baseline models across multiple real-world benchmark datasets.

\end{abstract}

\section{Introduction}

The detection of overlapping communities, where nodes may simultaneously belong to multiple densely-connected groups, is fundamental for analyzing complex networks. These structures naturally arise in social networks, biological systems, and information networks~\cite{1}. Traditional community detection identifies node clusters with stronger internal than external connections, but overlapping communities present unique challenges as they often exhibit richer inter-community links~\cite{1,2}. Understanding these overlaps improves structural insight and predictive accuracy. However, many existing methods struggle with high-dimensional node and community data, thus limiting detection performance. While some incorporate community-level features, their computational complexity hinders scalability on large networks.

Community detection methods can be divided into traditional and deep learning-based approaches~\cite{3}. Traditional models like AGM~\cite{4}, BIGCLAM~\cite{5}, and CESNA~\cite{6} use affiliation graphs to detect overlapping communities based on topology. BIGCLAM improves scalability by modeling affiliations as continuous variables, while CESNA incorporates node attributes to enhance accuracy. However, these methods are computationally intensive and struggle to capture the complex nonlinear relationships in real-world networks.

In contrast, Graph Convolutional Networks (GCNs)~\cite{3,7,8,9,10,11} excel at tasks like node classification and link prediction by learning low-dimensional embeddings that capture nonlinear patterns. NOCD~\cite{12} integrates the Bernoulli-Poisson model~\cite{5,13,14} with GCNs to learn community affiliations in an end-to-end manner, effectively utilizing node features but neglecting intrinsic community structures. UCoDe~\cite{15} adopts modularity~\cite{16} from a global perspective to infer community affiliations via GCNs. While effective on small- and medium-scale networks, its performance declines on larger graphs due to unrealistic assumptions about inter-community connectivity~\cite{17} and its tendency to overlook small communities. CDMG~\cite{9} replaces modularity with Markov stability, optimizing over Markov time to enhance detection quality. However, its high computational demands and sensitivity to time parameters limit scalability.

To address the limitations of traditional and GCN-based methods—such as limited structural utilization, high computational costs, and inadequate modeling of community structure—this paper proposes LQ-GCN, a local-perspective model for overlapping community detection. By integrating the Bernoulli-Poisson model, LQ-GCN jointly learns node embeddings and community affiliations from both topology and attributes. It incorporates local modularity to evaluate connectivity between communities and their neighbors, refining boundaries and enhancing the detection of densely connected substructures. The GCN architecture is further optimized for robustness on large-scale networks. Experiments on real-world datasets demonstrate that LQ-GCN achieves superior performance, with significantly higher normalized mutual information than baseline models. Ablation studies validate the contributions of local modularity and architectural enhancements to scalability and accuracy.


\begin{figure*}[htpb]
  \centering
  \includegraphics[width=\linewidth]{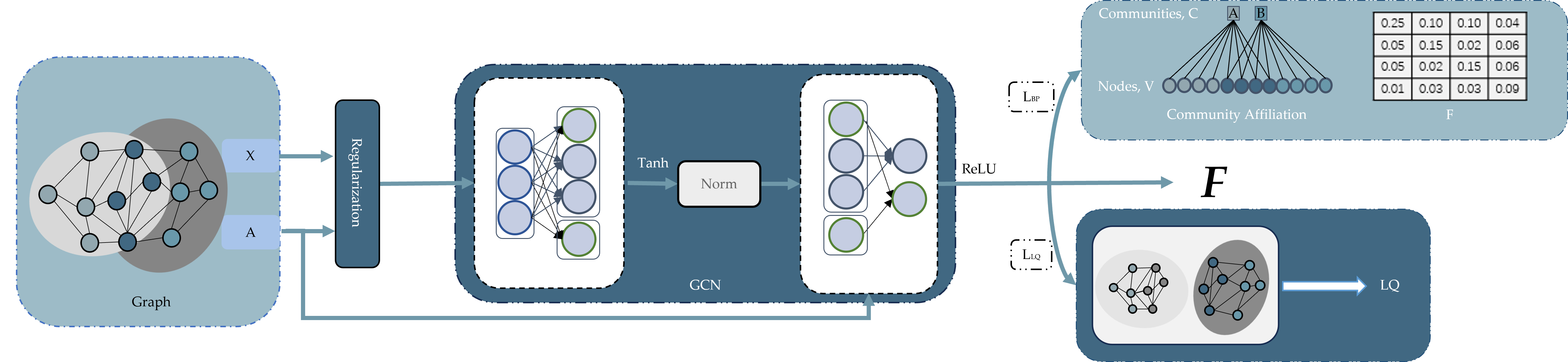}
  \caption{The proposed LQ-GCN Model Framework}
  \label{fig1}
\end{figure*}

\section{Methodology}

The proposed LQ-GCN model is tailored for efficient overlapping community detection in large-scale networks with three key components: (1) a GCN forms the backbone, processing the adjacency matrix $A$ and node attribute matrix $X$, with adaptive modifications to enhance node discriminability in large-scale settings; (2) a Bernoulli-Poisson (B-P) model is integrated into the representation learning process, enabling end-to-end community detection; and (3) local modularity~\cite{17} is incorporated into the loss function to guide the learning of accurate community memberships, thereby improving detection quality and stability.

As illustrated in Fig.~\ref{fig1}, the model begins by feeding the adjacency matrix $A$ and attribute matrix $X$ into a normalization module to generate a regularized matrix. The derived regularized matrix is then input into an improved GCN layer. The first convolutional layer employs a Tanh activation function, and its normalized output is passed to the second layer. The final optimization combines the B-P model with local modularity as the objective function, producing the node–community affiliation matrix $F$, which enables accurate and efficient detection of overlapping communities.

\subsection{GCN Architecture}
GCNs capture complex network relationships through a message-passing mechanism that propagates feature information across nodes. This process mirrors real-world information diffusion, which enhances the realism of inferred community structures. A single-layer GCN convolution is defined as:
\begin{equation}
H^{(l+1)} = \sigma\left( \tilde{D}^{-\frac{1}{2}} \tilde{A} \tilde{D}^{-\frac{1}{2}} H^{(l)} W^{(l)} \right)
\end{equation}
where $\sigma$ is the activation function;  
$\tilde{A} = A + I_N$ is the adjacency matrix with added self-loops;  
$\tilde{D}$ is the degree matrix of $\tilde{A}$;  
$H^{(l)}$ is the node feature matrix at layer $l$;  
and $W^{(l)}$ is the learnable weight matrix at layer $l$.  
The convolution operation involves iterative aggregation of information from neighboring nodes and the node itself, which results in progressive feature refinement.

To accommodate large-scale networks, LQ-GCN employs a two-layer convolutional architecture and modifies the original formulation as follows:
\begin{equation}
    F = ReLU\left(A \tanh\left(\bar{A}XW^{(1)} + XW^{(1)}\right)W^{(2)}\right)
\end{equation}
where $\bar{A} = I + D^{-\frac{1}{2}} A D^{-\frac{1}{2}}$ represents the normalized adjacency matrix, and $W^{(1)}$, $W^{(2)}$ are learnable parameters for the first and second layers, respectively. The hidden layer dimension is set to 128, while the output layer dimension corresponds to the number of target communities. The final output matrix $F$ represents node–community affiliations.

To mitigate overfitting, $L2$ regularization is applied to the model parameters, and dropout is used with a rate of 0.5. Normalization is performed in both convolutional layers. These techniques help prevent overfitting and oversmoothing issues in GCNs,.

\subsection{Loss Function}
To construct an end-to-end detection framework and enhance the quality of clustering results, we design a composite objective function comprising two components: $L_{BP}$ and $L_{LQ}$. The first term, $L_{BP}$, is based on the Bernoulli-Poisson (B-P) model and tailored for overlapping community detection. The second term, $L_{LQ}$, is inspired by local modularity $LQ$~\cite{17} and introduces a localized perspective to further refine the clustering performance. The following subsections detail the formulation and role of each loss component.

\subsubsection{Bernoulli-Poisson Loss $L_{BP}$}
The B-P model is a bipartite affiliation model designed to uncover overlapping communities by learning a community affiliation matrix $F$ that approximates the observed adjacency matrix $A$:
\begin{equation}
    A_{ij} \sim Bernoulli\left(1 - \exp\left(-F_i F_j^{T}\right)\right)
\end{equation}

In this model, the probability of an edge between nodes $V_{i}$ and $V_{j}$ is positively correlated with the number of communities they share. Here, $V$ represents the set of vertices, and $E$ denotes the set of edges in the graph. Each entry $F_{ij}$ in the matrix F denotes the likelihood that node $V_{i}$ belongs to community $C_{j}$. If $F_{ij}$ exceeds a predefined threshold $p$, node $V_{i}$ is assigned to community $C_{j}$. A node may belong to multiple communities if it has multiple high-affiliation values, which thus enables the detection of overlapping communities. The loss function $L_{BP}$ to be minimized is defined as:
\begin{equation}
    L_{BP} = \sum_{(i,j) \in E} \log\left(1 - \exp\left(-F_i F_j^{T}\right)\right) - \sum_{(i,j) \notin E} F_i F_j^{T}
\end{equation}
Here, $F_i F_j^T$ represents the inner product between rows $i$ and $j$ of the matrix $F$, which estimates the connection strength. Given that real-world networks are typically sparse, the number of non-edges far exceeds that of actual edges,  
making the second summation dominate the loss. To address this imbalance, the final weighted formulation becomes:
\begin{align}
    L_{BP} = -{\textbf{E}}_{(i,j) \sim {P}_{E}} 
    \left[\log\left(1 - \exp\left(-F_i F_j^{T}\right)\right)\right] \notag + {\textbf{E}}_{(i,j) \sim {P}_{N}} \left[F_i F_j^{T}\right]
\end{align}
Here, $\textbf{E}[\cdot]$ denotes the expectation, with $P_{E}$ and $P_{N}$ denoting uniform distributions over connected and non-connected node pairs, respectively.

\subsubsection{Local Modularity Loss $L_{LQ}$}
Modularity $Q$, introduced by Newman~\cite{16}, quantifies the quality of a graph clustering $C$ on graph $G$:
\begin{equation}
    Q(G, C) = \frac{1}{2|{E}|} \sum_{s}^{K} \sum_{ij} \left(A_{ij} - \frac{d_i d_j}{2|{E}|} \right) C_{is} C_{js}
\end{equation}
where $d_i$ represents the degree of node $i$ (the number of edges connected to the node) and $K$ denotes the number of communities. While higher modularity scores indicate better clustering quality, the assumption of uniform inter-community connection probabilities does not hold in real-world networks. Consequently, models like UCoDe~\cite{15}, which incorporate $Q$ into their loss function, struggle with scalability and accuracy on large-scale graphs.

To overcome these limitations, Local Modularity $LQ$~\cite{17} is proposed to evaluate connectivity between a community and its neighboring communities instead of the entire network. The focus on locality enhances clustering granularity and reduces global dependency. The local modularity score for community $C$ is defined as:
\begin{equation}
    LQ(G, C) = \frac{1}{2|E|} \sum_{s}^{K} \sum_{ij} \left( \frac{A_{ij}}{|L_{iN}|} - \frac{d_i d_j}{2|L_{iN}|} \right) F_{is} F_{js}
\end{equation}
where $L_{iN}$ represents the neighborhood size of node $i$ (number of edges connecting to community $C_i$). By defining a local modularity matrix $B$ with $B_{ij} = \frac{A_{ij}}{|E|} - \frac{d_i d_j}{|E|}$, the Equation (7) can be simplified as:
\begin{equation}
    LQ(Q, C) = \frac{1}{4|E|} \, Tr(SC^{T}BC)
\end{equation}
Here, $S \in [0,|E|]^{k\times k}$ is a matrix where $S_{ii}=|E|/|L_{iN}|$, representing the ratio of total edges to the number of edges between community $C_{i}$ and its neighbors. The resulting matrix $LQ_{M}=SC^TBC$ captures both intra- and inter-community local modularity, with diagonal elements representing intra-community modularity and off-diagonal elements representing inter-community similarity.

To enhance clustering quality, we aim to maximize diagonal elements while minimizing off-diagonal ones. The loss is then defined as:
\begin{equation}
    L_{LQ} = -\frac{1}{2k} \sum_{s=1}^{k} \Big( y_s \log\left(dg(LQ_{M})_{s}\right) \notag  + (1 - y_{k+s}) \log\left(1 - dg(\theta(LQ_{M})_{s})\right) \Big)
\end{equation}
where $dg(\cdot)$ extracts diagonal entries, $\theta(\cdot)$ performs exclusive comparisons between communities, and $y_{s} \in \{0,1\}$ is the target distribution which indicates whether a sample corresponds to intra-community ($y_{s}=1$) or inter-community ($y_{s}=0$) modularity. This design ensures that while clustering similar nodes enhances intra-community coherence, the distinction between different nodes is maintained, thereby improving community detection accuracy.


Finally, the total loss function for LQ-GCN is defined as $L = \alpha L_{BP} + \beta L_{LQ}$, where $\alpha$ and $\beta$ are balancing coefficients. The model is trained using the Adam optimizer.

\subsection{LQ-GCN Algorithm}

Algorithm~\ref{algorithm1} summarizes the detailed procedure of the proposed LQ-GCN model. In Lines 1–3, the input adjacency matrix is preprocessed to obtain the normalized adjacency matrix $\bar{A}$ and augmented matrix $\tilde{A}$, which are then combined with the attribute matrix $X$ as inputs to the first and second convolutional layers. Weight matrices $W^{(1)}$ and $W^{(2)}$ are initialized accordingly. Lines 4–15 define an early stopping strategy to monitor loss convergence and control the number of iterations, thereby reducing computational overhead. Once the loss stabilizes for more than 30 iterations, the local modularity loss $L_{LQ}$ is introduced to refine community assignments. Training is terminated if the loss fails to improve for more than 80 iterations. In Line 16, after two-layer convolution and ReLU activation, each entry $F_{ij}$ in the output matrix represents the probability that node $V_i$ belongs to community $C_j$. If $F_{ij}$ exceeds a predefined threshold $\text{thresh}$, node $V_i$ is assigned to community $C_j$. The resulting matrix $F$ constitutes the final community affiliation output.

\begin{algorithm}
\caption{LQ-GCN}
\label{algorithm1}
\textbf{Input:} Adjacency matrix $A$, attribute matrix $X$, number of communities $K$, number of iterations $Itr$ \\
\textbf{Output:} Community affiliation matrix $F$
\begin{algorithmic}[1]
\State Apply L2 normalization to $A$ and $X$
\State Compute normalized adjacency matrix $\tilde{A} = I + D^{-1/2} A D^{-1/2}$ and augmented matrix $\tilde{A} = A + I_N$
\State Initialize weight matrices $W^{(1)}$ and $W^{(2)}$ using Xavier initialization
\State Initialize \texttt{early\_stopping} to track consecutive non-decreasing loss; reset when a new minimum is observed
\For{$idx = 1$ to $Itr$}
    \State Compute community affiliation matrix $F$ by (4)
    \State Calculate $L_{BP}$ by (5) as the initial objective function
    \If{$early\_stopping > 30$}
        \State Include $L_{LQ}$ by (10) to improve clustering
    \EndIf
    \If{$early\_stopping > 80$}
        \State \textbf{break}
    \EndIf
    \State Update model parameters using the Adam optimizer
\EndFor
\If{$F_{ij} > \text{thresh}$}
    \State Set $F_{ij} = 1$
\EndIf
\State \Return $F$
\end{algorithmic}
\end{algorithm}

\section{Experiments and Evaluation}
\subsection{Datasets and Evaluation Metrics}
To evaluate model performance, we conduct experiments on six publicly available real-world datasets~\cite{12}, encompassing networks of varying scales, with node counts ranging from 170 to 35,409. The datasets include three Facebook social networks and three co-authorship networks from the Microsoft Academic Graph, covering the fields of chemistry, computer science, and engineering.

The Facebook datasets, ranging from 170 to 792 nodes, represent small-scale social networks where node features correspond to users’ personal attributes. To assess model performance on large-scale networks, three co-authorship networks are used, with node counts between 14,957 and 35,409. In these datasets, nodes represent papers and node attributes are extracted from their associated keywords. Tab.~\ref{tab1} summarizes key statistics for each dataset, including the number of nodes $N$, edges $|E|$, ground-truth communities $K$, and the dimensionality of node attributes $|X|$.

\begin{table}[htbp]
\centering
\caption{Real-world datasets for community detection}
\resizebox{0.45\columnwidth}{!}{%
\begin{tabular}{lccccc}
\toprule
\textbf{Dataset} & \textbf{$N$} & \textbf{$|E|$} & \textbf{$K$} & \textbf{$|X|$} \\
\midrule
Facebook 348          & 227    & 6384     & 14 & 21   \\
Facebook 686          & 170    & 3312     & 14 & 9    \\
Facebook 1684         & 792    & 28048    & 17 & 15   \\
Engineering           & 14927  & 98610    & 16 & 4839 \\
Computer Science      & 21597  & 193500   & 18 & 7793 \\
Chemistry             & 35409  & 314716   & 14 & 4877 \\
\bottomrule
\end{tabular}%
}
\label{tab1}
\end{table}

To evaluate the performance of overlapping community detection, we adopt Overlapping Normalized Mutual Information (ONMI)~\cite{18} and Recall as the primary evaluation metrics. These metrics quantify the statistical similarity between the detected clusters and the ground-truth labels, thereby reflecting clustering quality.

ONMI ranges from 0 to 1, where 0 indicates complete dissimilarity and 1 indicates perfect agreement between the predicted and true communities. A higher ONMI value suggests a better match to the true community structure. ONMI is defined as:
$ONMI(Y, C) = \frac{I(Y, C)}{\max(H(Y) + H(C))}$, where $Y$ denotes the ground-truth community assignments, $C$ denotes the predicted clustering results, $H$ is the entropy function, and $I(Y, C)$ represents the mutual information between $Y$ and $C$, calculated as $I(Y, C) = \tfrac{1}{2} \left[ H(Y) - H(Y \mid C) + H(C) - H(C \mid Y) \right]$. ONMI is preferred over traditional similarity metrics due to its intuitive interpretation and robustness. Its normalization procedure eliminates the influence of network scale and class number, making it particularly well-suited for evaluating overlapping communities in complex networks.

Recall assesses the model’s ability to correctly identify true community members. In the context of overlapping community detection, recall measures whether nodes are accurately assigned to all communities to which they actually belong. We adopt the average recall of best-matched clusters as defined in~\cite{5}, computed as $\text{Recall}(C^*, \hat{C}) = \frac{1}{|C^*|} \sum_{C_i \in C^*} Rc\left(C_i, \hat{C}_{g(i)}\right)$, where $C^*$ is the set of ground-truth communities, $\hat{C}$ is the set of predicted communities, and each community is defined by its set of member nodes. For each ground-truth community $C_{i}$, recall is computed by matching it to the predicted community $g$ with the largest node overlap.

\subsection{Comparative Experiments}
In this section, we compare the proposed LQ-GCN with several state-of-the-art methods, including BIGCLAM~\cite{5} and its extension CESNA~\cite{S1}, as well as GCN-based models such as NOCD~\cite{12}, UCoDe~\cite{15}, and CDMG~\cite{9}. Three input strategies denoted by different suffixes are used for all models: ``-X'' uses both adjacency matrix $A$ and node attribute matrix $X$, ``-G'' uses only $A$, and ``-U'' concatenates $A$ and $X$ as input. A brief overview of the baseline methods is provided below:
\begin{itemize}
    \item BIGCLAM~\cite{5}: A community affiliation model that detects overlapping communities by maximizing the likelihood of the observed graph. Unlike traditional NMF, it (1) directly models the graph structure, (2) allows overlapping memberships, and (3) is optimized for large, sparse networks with complexity proportional to the number of edges.
    \item CESNA~\cite{S1}:An extension of BIGCLAM that integrates network structure and node attributes, assuming both are generated from underlying communities. While more computationally demanding, it performs well on sparse networks with rich attribute data.
    \item NOCD~\cite{12}: Among the first to apply GNNs to overlapping community detection, it jointly learns node embeddings and community affiliations in a scalable end-to-end framework, achieving strong performance on large graphs.
    \item UCoDe~\cite{15}: A unified framework for detecting both overlapping and non-overlapping communities. Unlike NOCD’s Bernoulli-Poisson loss, it employs a modularity-based loss, extending traditional modularity optimization to support overlaps, enhancing interpretability and flexibility.
    \item CDMG~\cite{9}: Combines GCNs with Markov Stability to assess the dynamic consistency of communities. It improves clustering by aligning embeddings with topology, but its reliance on partitioning large networks limits performance on full-scale graphs.
\end{itemize}

\begin{table*}[htbp]
\centering
\caption{The ONMI and Recall percentage scores of overlapping community detection}
\resizebox{\linewidth}{!}{%
\begin{tabular}{lcccccccccccc}
\toprule
\textbf{Method}
& \multicolumn{2}{c}{\textbf{Facebook-686}} 
& \multicolumn{2}{c}{\textbf{Facebook-348}} 
& \multicolumn{2}{c}{\textbf{Facebook-1684}} 
& \multicolumn{2}{c}{\textbf{Engineering}} 
& \multicolumn{2}{c}{\textbf{Computer Science}} 
& \multicolumn{2}{c}{\textbf{Chemistry}} \\
\cmidrule(lr){2-3} \cmidrule(lr){4-5} \cmidrule(lr){6-7} \cmidrule(lr){8-9} \cmidrule(lr){10-11} \cmidrule(lr){12-13}
& ONMI & Recall 
& ONMI & Recall 
& ONMI & Recall 
& ONMI & Recall 
& ONMI & Recall 
& ONMI & Recall \\
\midrule
BIGCLAM     & 13.8 & 32.6 & 26.0 & 34.0 & 32.7 & 37.5 & 7.9  & 18.5 & 0.0  & 26.7 & 0.0  & 30.2 \\
CESNA       & 13.3 & 33.3 & 29.4 & 31.4 & 28.0 & 30.2 & 24.3 & 21.3 & 33.8 & 22.8 & 23.3 & 26.8 \\
NOCD-G      & 17.9 & 28.3 & 33.2 & 40.0 & 36.0 & 49.6 & 18.4 & 48.8 & 34.2 & 46.7 & 22.6 & 43.6 \\
NOCD-X      & 18.5 & 25.3 & 27.9 & 46.0 & 30.0 & 44.6 & 39.1 & 46.2 & 50.2 & 42.3 & 45.3 & 46.5 \\
CDMG        & 13.4 & 27.3 & 28.5 & 47.6 & 29.4 & 42.4 & 4.1  & 24.1 & 3.4  & 23.4 & --   & --  \\
UCoDe-U     & 14.1 & 23.1 & 24.4 & \textbf{77.0} & 31.3 & 50.5 & 1.4  & 39.9 & 21.9 & 33.8 & 19.6 & 33.6 \\
UCoDe-X     & \textbf{23.8} & \textbf{56.0} & 27.2 & 44.1 & 21.5 & \textbf{64.5} & 1.8  & 35.2 & 20.8 & 34.5 & 25.9 & 38.8 \\
LQ-GCN-X    & 20.3 & 31.1 & 32.6 & 48.5 & 27.2 & 47.2 & \textbf{42.1} & \textbf{56.4} & \textbf{53.8} & \textbf{60.8} & \textbf{47.1} & \textbf{59.1} \\
LQ-GCN-G    & 20.8 & 29.5 & \textbf{35.8} & 52.6 & \textbf{44.1} & 59.1 & 21.9 & 49.2 & 39.1 & 54.7 & 27.1 & 52.6 \\
LQ-GCN-U    & 19.5 & 26.2 & 31.3 & 51.3 & 35.9 & 55.9 & 26.6 & 48.4 & 38.5 & 56.1 & 34.5 & 56.4 \\
\bottomrule
\end{tabular}%
}
\label{tab2+3combined}
\end{table*}

Tab.~\ref{tab2+3combined} presents the ONMI and Recall scores across all datasets. The proposed LQ-GCN outperforms baseline methods on both small-scale (Facebook) and large-scale (co-authorship) networks. For example, LQ-GCN-X improves ONMI on Facebook-348 from 27.9\% (NOCD-X) to 32.6\% and Recall from 46.0\% to 48.5\%. On Facebook-686, ONMI increases from 18.5\% to 20.3\% and Recall from 25.3\% to 31.1\%. LQ-GCN-G achieves the highest ONMI on Facebook-348 and Facebook-1684, while LQ-GCN-X obtains the best Recall on the large-scale Engineering, Computer Science, and Chemistry networks. Interestingly, on small-scale networks, models using only adjacency matrices $(G)$ often outperform those using node attributes $(X)$, suggesting that topological structure is more informative than attribute features when node diversity is limited.

On large-scale datasets, LQ-GCN-X consistently achieves the highest ONMI and Recall. For instance, on the Computer Science dataset, it surpasses NOCD by 3.8\%, UCoDe by 33\%, and CDMG by 50.4\% in ONMI, while Recall improves by 18.5\% and 25.8\% over NOCD and UCoDe, respectively. These results confirm LQ-GCN’s strong scalability and adaptability. Although UCoDe performs well on small datasets, its modularity assumption limits performance on large-scale networks. Similarly, CDMG partitions large graphs into subgraphs, achieving good results locally but underperforming at scale.

\begin{figure}[htbp]
    \centering
    \begin{subfigure}{0.8\linewidth}
        \centering
        \includegraphics[width=\linewidth]{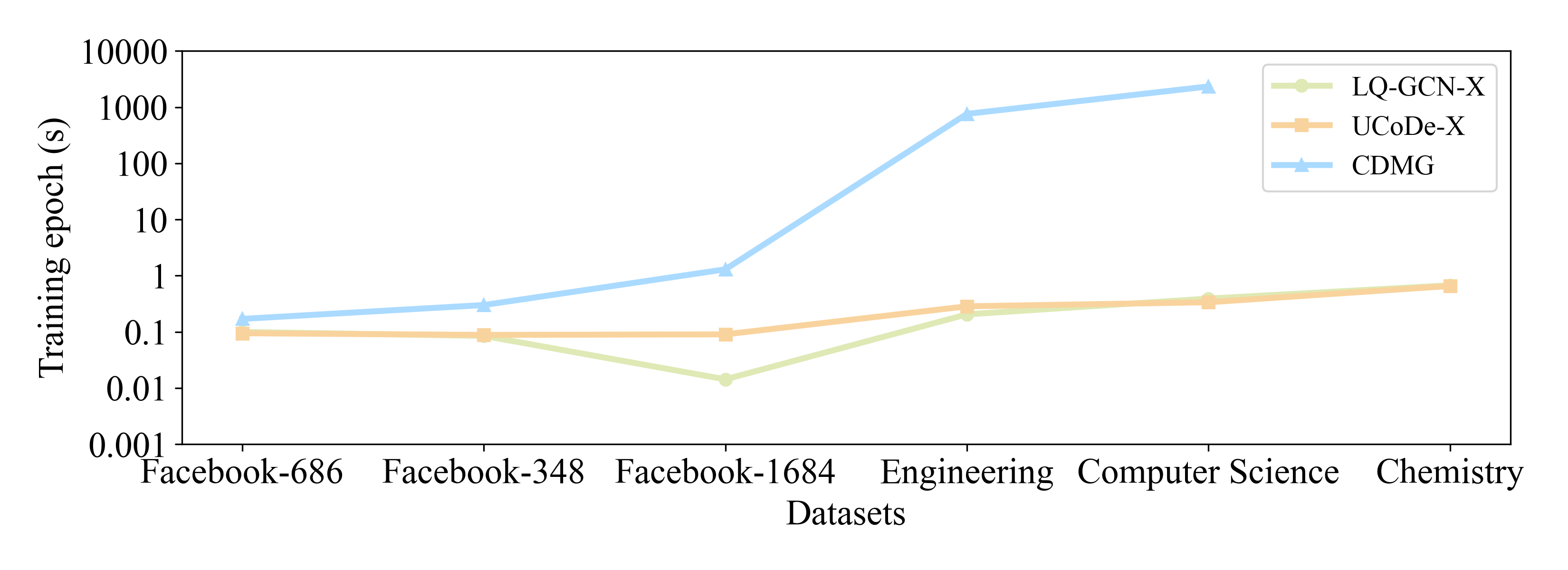}
        \caption{Training epoch (s)}
    \end{subfigure}
    \hfill
    \begin{subfigure}{0.8\linewidth}
        \centering
        \includegraphics[width=\linewidth]{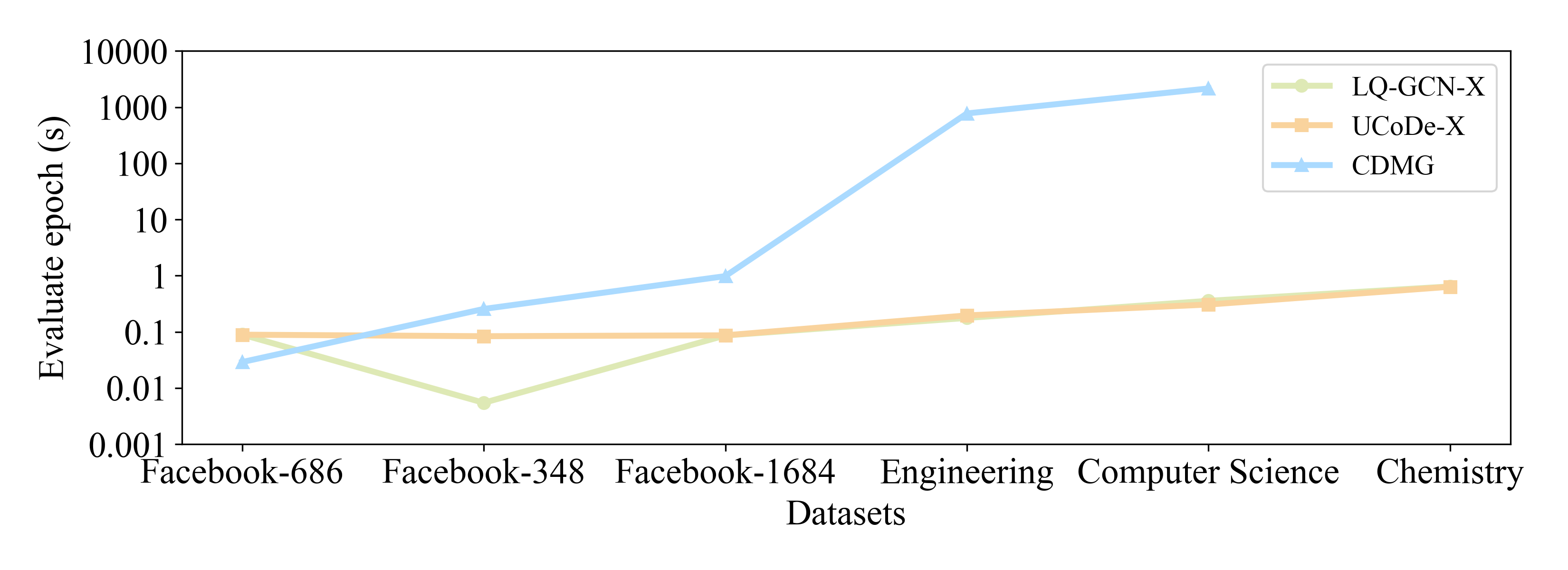}
        \caption{Evaluate epoch (s)}
    \end{subfigure}
    \caption{The runtime of community detection algorithms.}
    \label{fig2}
\end{figure}

We also compare training and inference time for LQ-GCN-X, UCoDe-X, and CDMG. As shown in Fig.~\ref{fig2}, LQ-GCN requires slightly more time than UCoDe due to the added cost of computing local modularity. Nevertheless, LQ-GCN significantly outperforms CDMG in both efficiency and ONMI, further validating its effectiveness for large-scale community detection. In summary, while GCN-based models like UCoDe and NOCD excel on small networks, LQ-GCN exhibits superior robustness and generalization, particularly in large-scale and attribute-scarce scenarios.

\subsection{Ablation Study}
To assess the contribution of local modularity and the modified GCN architecture, we conduct ablation experiments on all datasets listed in Tab.~\ref{tab1}. We define LQ-GCN-LX as the model without local modularity, and LQ-GCN-GX as the model without modified convolution layers. All experiments are run 50 times using a threshold $p$=0.5 and a hidden layer size of 128.

\subsubsection{Effect of Local Modularity}
To evaluate the impact of local modularity, we remove $L_{LQ}$ from loss $L$. As shown in Tab.~\ref{tab4+5combined}, LQ-GCN-X consistently outperforms LQ-GCN-LX across datasets, with especially large improvements on the Computer Science dataset (ONMI ↑15.9\%) and Chemistry dataset (Recall ↑15.2\%). While gains are smaller on some small datasets like Facebook-348, the improvement in Recall is substantial. This confirms the effectiveness of incorporating local modularity despite added computational cost.

\begin{table}[htbp]
\centering
\caption{Performance comparison of different model structures}
\resizebox{0.7\linewidth}{!}{%
\begin{tabular}{c c c c c c}
\toprule
\textbf{Dataset} & \textbf{Metric} & \textbf{LQ-GCN-X} & \textbf{LQ-GCN-LX} & \textbf{LQ-GCN-GX} \\
\midrule
\multirow{2}{*}{Facebook-348}     & ONMI   & \textbf{0.216} & 0.172 & 0.172 \\
                                  & Recall & \textbf{0.484} & 0.392 & 0.412 \\
\multirow{2}{*}{Facebook-686}     & ONMI   & \textbf{0.393} & 0.324 & 0.321 \\
                                  & Recall & \textbf{0.313} & 0.236 & 0.265 \\
\multirow{2}{*}{Facebook-1684}    & ONMI   & \textbf{0.287} & 0.216 & 0.217 \\
                                  & Recall & \textbf{0.472} & 0.395 & 0.391 \\
\multirow{2}{*}{Engineering}      & ONMI   & \textbf{0.416} & 0.394 & 0.396 \\
                                  & Recall & \textbf{0.564} & 0.481 & 0.466 \\
\multirow{2}{*}{Computer Science} & ONMI   & \textbf{0.541} & 0.382 & 0.384 \\
                                  & Recall & \textbf{0.608} & 0.512 & 0.476 \\
\multirow{2}{*}{Chemistry}        & ONMI   & \textbf{0.468} & 0.367 & 0.365 \\
                                  & Recall & \textbf{0.591} & 0.439 & 0.423 \\
\bottomrule
\end{tabular}%
}
\label{tab4+5combined}
\end{table}

\subsubsection{Effect of Modified Convolution}
To examine the impact of convolutional layer improvements, we define LQ-GCN-GX using two unmodified GCN layers with identical activation functions:

\begin{equation}
    F = ReLU\left( \bar{A} \tanh\left( \bar{A}H^{(l)}W^{(1)} \right) W^{(2)} \right)
\end{equation}


Tab.~\ref{tab4+5combined} compares LQ-GCN-X with LQ-GCN-GX. On small datasets, improvements are modest (ONMI and Recall gains $<$ 10\%), but on large networks, LQ-GCN-X significantly outperforms LQ-GCN-GX (e.g., ONMI ↑15.7\%, Recall ↑13.2\%). Although modified layers introduce additional computational cost, they enhance the utilization of node attributes and lead to more accurate community detection. Thus, LQ-GCN-GX is recommended for small graphs and LQ-GCN-X on large networks.

\subsection{Parameter Sensitivity Analysis}
LQ-GCN generates a soft affiliation matrix $F$ and assigns node $V_i$ to community $C_j$ if $F_{ij}$ $>$ $p$. A lower $p$ may result in excessive community assignment, while a higher $p$ may miss valid overlaps. We run the experiments 20 times with 1000 iterations for each setting and report the average performance.

\begin{figure}[htbp]
    \centering
    \includegraphics[width=\linewidth]{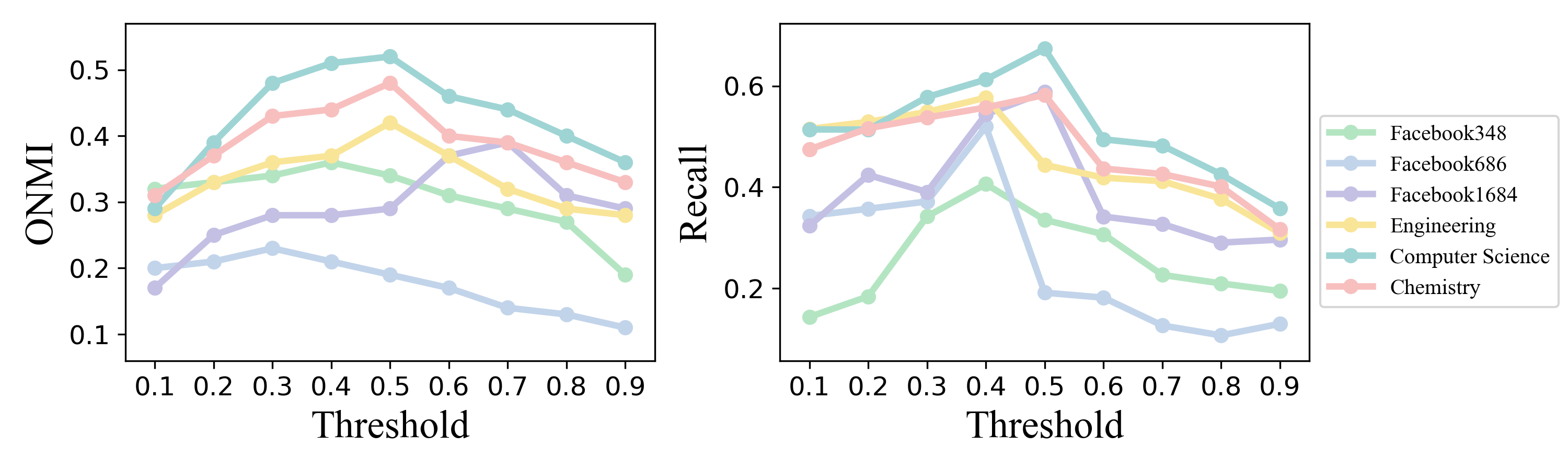}
    \caption{The influence of the threshold.}
    \label{fig3}
\end{figure}

As shown in Fig.~\ref{fig3}, the model achieves the best overall performance when $p=0.5$. Although some datasets (e.g., Facebook-686, Facebook-1684) perform slightly better with $p=0.4$ or $p=0.6$, the general trend supports $p=0.5$ as the optimal setting for stable and effective overlapping community detection.

\section{Conclusion}
This paper proposes LQ-GCN, an overlapping community detection method based on local modularity, addressing the limitations of existing GCN-based approaches on large-scale networks due to their insufficient modeling of community structures. By integrating local modularity into the loss function and enhancing the GCN architecture, LQ-GCN significantly improves detection performance. Extensive experiments on public datasets demonstrate that LQ-GCN improves ONMI by up to 33\% and improves Recall by up to 26.3\% over the advanced UCoDe model, showing strong performance advantages. Ablation studies further validate the crucial roles of local modularity and the modified convolutional layers in enhancing accuracy and robustness. Future work will focus on optimizing LQ-GCN and extending it to more complex and heterogeneous network scenarios to further enhance its practical applicability.

\printbibliography

\end{document}